# RELATIONSHIP BETWEEN DIFFUSION, SELFDIFFUSION AND VISCOSITY

**I. Avramov**

*Institute of Physical Chemistry, Bulgarian Academy of Sciences,*

*1113 Sofia, Bulgaria,*

*E-mail avramov@ipc.bas.bg*

## ABSTRACT

We investigate the experimental limits of validity of the Stokes-Einstein equation. There is an important difference between diffusion and self-diffusion. There are experimental evidences, that in the case of self-diffusion the product $D\eta /T$ is constant and the equation is still valid. However, comparison of existing experimental data on viscosity $\eta$ and diffusion coefficients $D$ of small, fast moving ions unambiguously show that the product $D\eta /T$ depends strongly on temperature $T$. The temperature dependence of diffusion coefficient declines from that of viscosity. Therefore, the Stokes-Einstein equation is not valid in this case.

## 1. Introduction

According to the famous Stokes-Einstein equation [1]

$$D = \frac{kT}{Br\eta} \tag{1}$$

diffusion coefficient $D$ is related to viscosity $\eta$ of the system through the radius $r$ of the moving particle. The absolute rate theory [2-4] predicts for the numerical coefficient value of $B= 2$ while the Stokes formula gives $B= 6\pi$. As soon as Eq.(1) was derived for hydrodynamic motion, it is reasonable to test whether it is applicable to atomic sized [5] particles. This test is important for the diffusion of molecules themselves. It is problematic for atoms and ions smaller the main building units of the structure of glassforming melt. If Eq.(1) is valid, the product of diffusion coefficient $D$, viscosity $\eta$ and temperature, $D\eta /T$, should be constant. Meanwhile, in Arrhenius coordinates (*lg(D/T) vs 1/T*), data on diffusion coefficient give quite a straight line. On the other hand, viscosity is known to be quite "*fragile"* (see for instance data in [6,7] )) and declines considerably from a straight line.

The diffusion coefficients, respectively of the *dc* conductivity due to the mobile cations [8–14], is determined by the decoupling of motions of the relatively weakly bounded cations, such as $Li^+$, $Na^+$ , and $Ag^+$, from the motions of host network formers, like *Si, B, Ge,* bounded to oxygen. The activation energy for the *dc* conductivity, taken from the slope of the Arrhenius plots, was examined and several models were developed to correlate it with the underlying short and intermediate range order of the glass. Anderson and Stuart [15], McElfresch and Howitt [16], and Elliott [17] correlate the activation barriers to short range structures of the glass, dividing the energy barrier into coulombic and strain energy parts. Ingram et al. [18] and Greaves et al. [19], relate the barriers to the long-range connectivity of the sites requisite for long range *dc* conduction. Ravaine and Souquet [20] and Souquet and Perera [21] describe the energy barriers in terms of the thermodynamics of dissolution, where the dissociation of the cation from its counter anion is required to become a mobile cation. In all of these models, however, simple activated rate theory was used to describe the temperature dependence of the process. The non-Arrhenius behavior was found in Refs. [7, 22, 23].

## 2. Experimental test of limits of validity os Stoekes-Einstein equation

Here, we demonstrate that the lower limit is the self-diffusion coefficient, i.e. the smallest particles for which this equation holds are the main building units of the systems. Smaller and faster moving particles does not follow it.

According to Eq.(1) the temperature dependencies of $lg\eta$ and of *lg(D/T)* should identical shifted by constant. An illustration of this is given in Fig.1 where the temperature dependence of viscosity (black points and left scale) of $Na_2O\ 2SiO_2$ glass according to Refs.[24,25] is represented against reciprocal temperature. Experimental results [24,25] on self-diffusion coefficient of $SiO_2$ are shown in the same figure (open points and right scale). It is seen that data go together reasonably well. This is an indication that, for self diffusion, Eq.(1) is still valid. The product $\lg\left(\frac{D\eta}{T}\right) = -12.2 \pm 0.1$ is in reasonable agreement with the expected $\lg\left(\frac{k}{Bd_o}\right) \approx -13$ value.

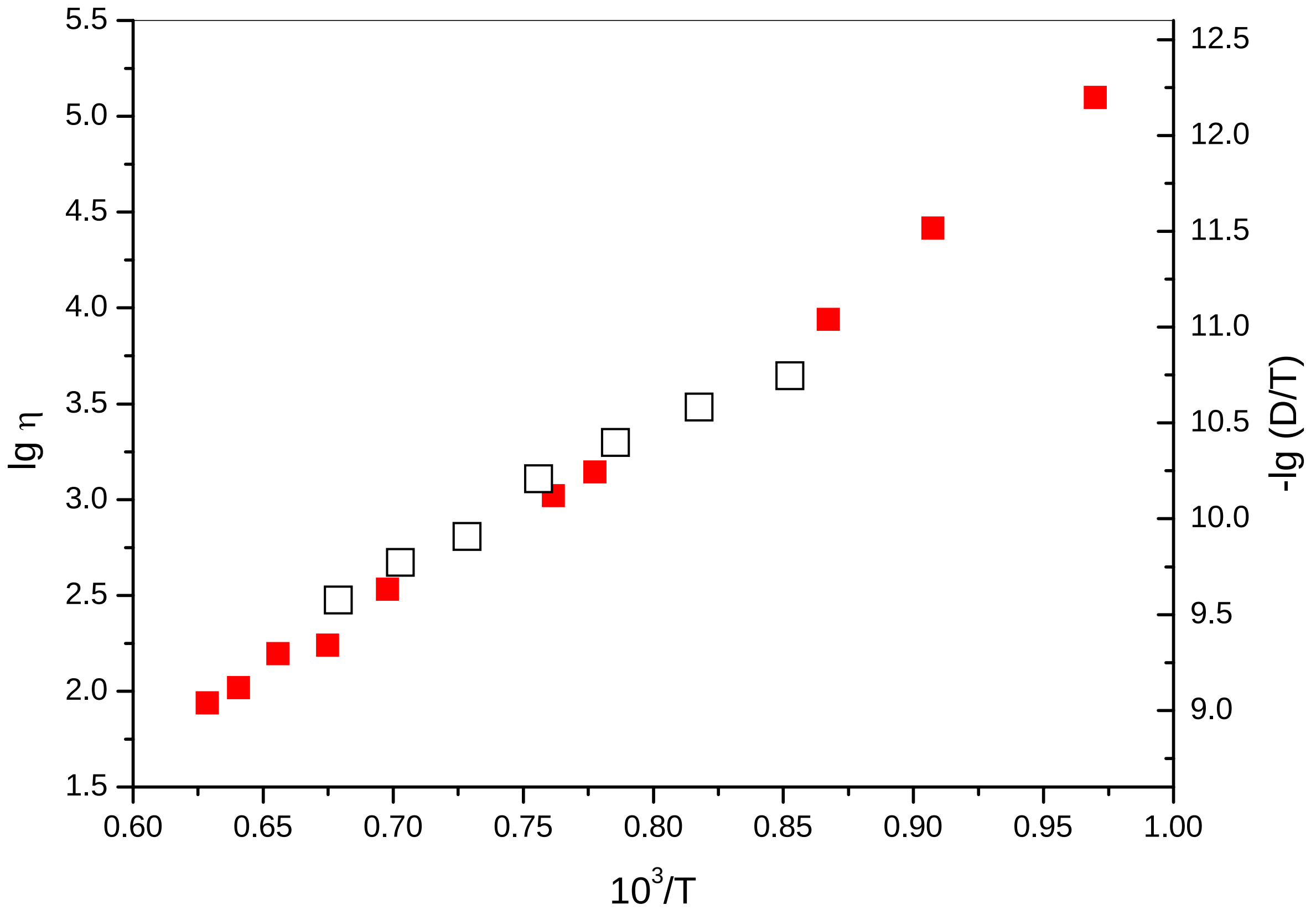

***Fig.1.*** *Arrhenius plot of viscosity, in [dPa.s] (■ points and left scale) and self-diffusion coefficient, in [cm²/s.K] (□ points and right scale) of $SiO_2$ in $Na_2O\ 2SiO_2$ glass according to experimental data published in Refs.[30,31].*

The situation could be different if we compare viscosity to diffusion coefficient of smaller and faster moving ions. Figure 2 gives in Arrhenius coordinates data [24,26,27] viscosity of pure $SiO_2$ (open circles and right scale) together with diffusion coefficient (*lg(D/T)*) of $Na^+$ (□ points); $K^+$ (▲ points); $Rb^+$ (Δ points) and $Cs^+$ (■ points). It is readily seen that viscosity has behavior completely different from that of the diffusion coefficients. The diffusion activation energy increases with ion radius [30, 32-34] as demonstrated on Fig.3. According to Eq.(1) it is expected that the product $\lg\left(\frac{D\eta}{T}\right) = Const - \lg d_o$ should decrease with logarithm of the size of the particle. The viscosity and diffusion curves overlap only at 1250 K, where we find that the product $\lg\left(\frac{D\eta}{T}\right)$ decrease much faster. From Fig.4 it is seen that the decrease is even faster than linear.

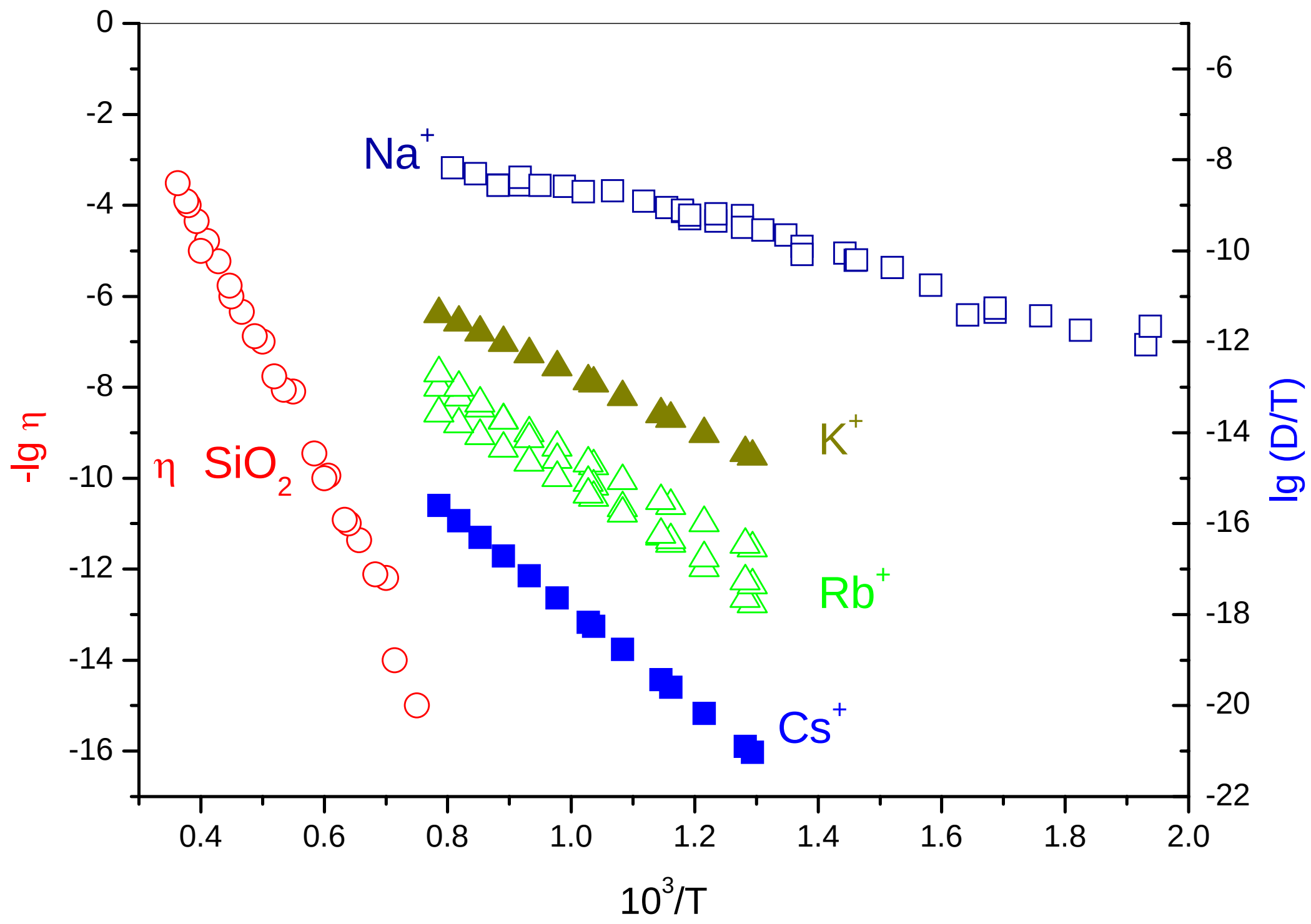

***Fig.2.*** *Arrhenius plot of viscosity of pure* $SiO_2$ *(o points and left scale) and diffusion coefficients (right scale) of* $Na^+$ *(□ points);* $K^+$ *(▲ points);* $Rb^+$ *(Δ points) and* $Cs^+$ *(■ points) according to experimental data published in Refs.[30,32,33].*

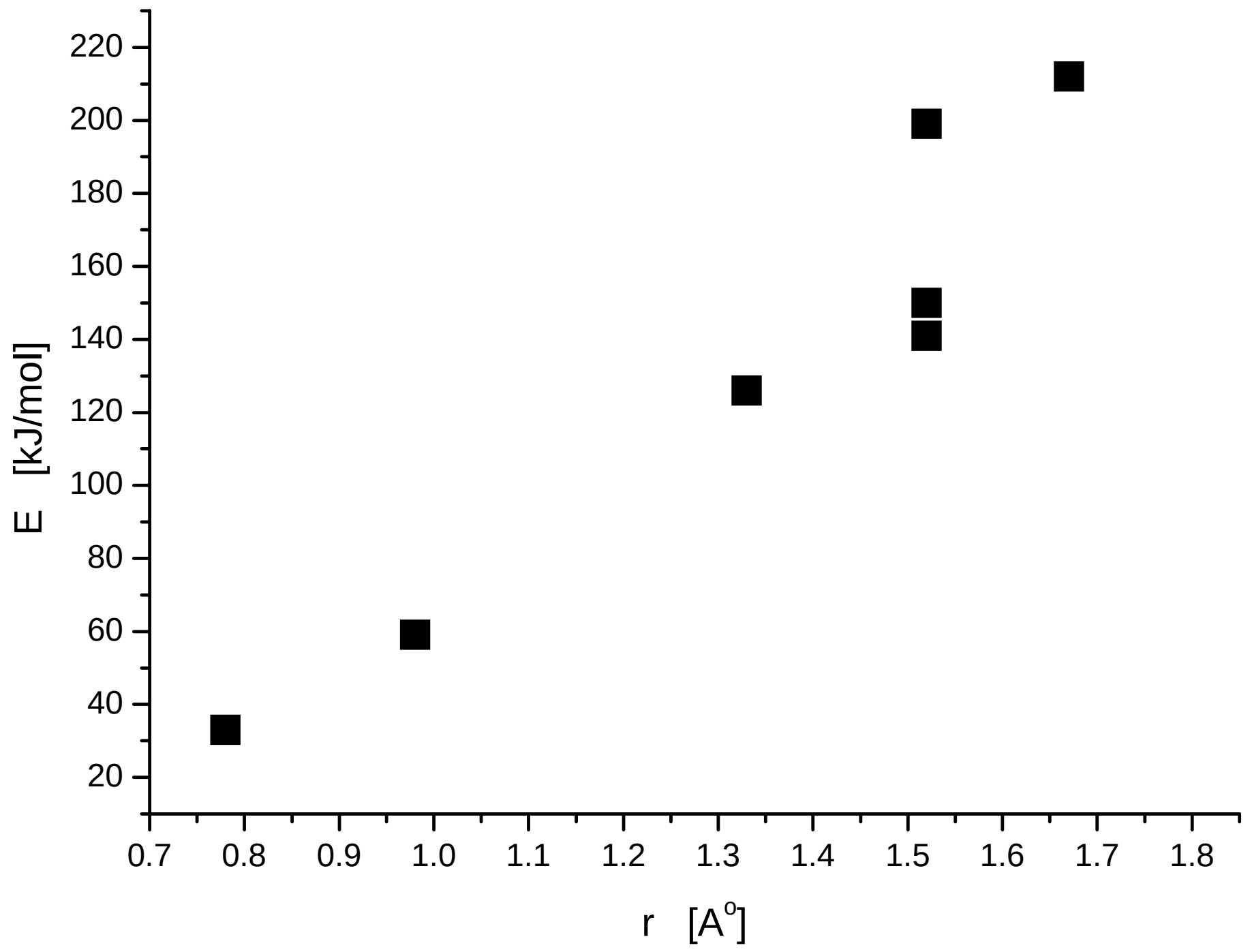

***Fig.3.*** *Diffusion activation energy versus ion radius according to experimental data published in Refs.[30,32-34].*

To widen the range of studied substances, we compared the experimental data [24] on viscosity and diffusion coefficient of *PbO* in $PbO\ 2B_2O_3$. Figure 5 gives the dependence of the product $\lg\left(\frac{D\eta}{T}\right)$ on temperature. Instead of being constant, the product decreases almost linearly with temperature. The value is dropping about thousand times.

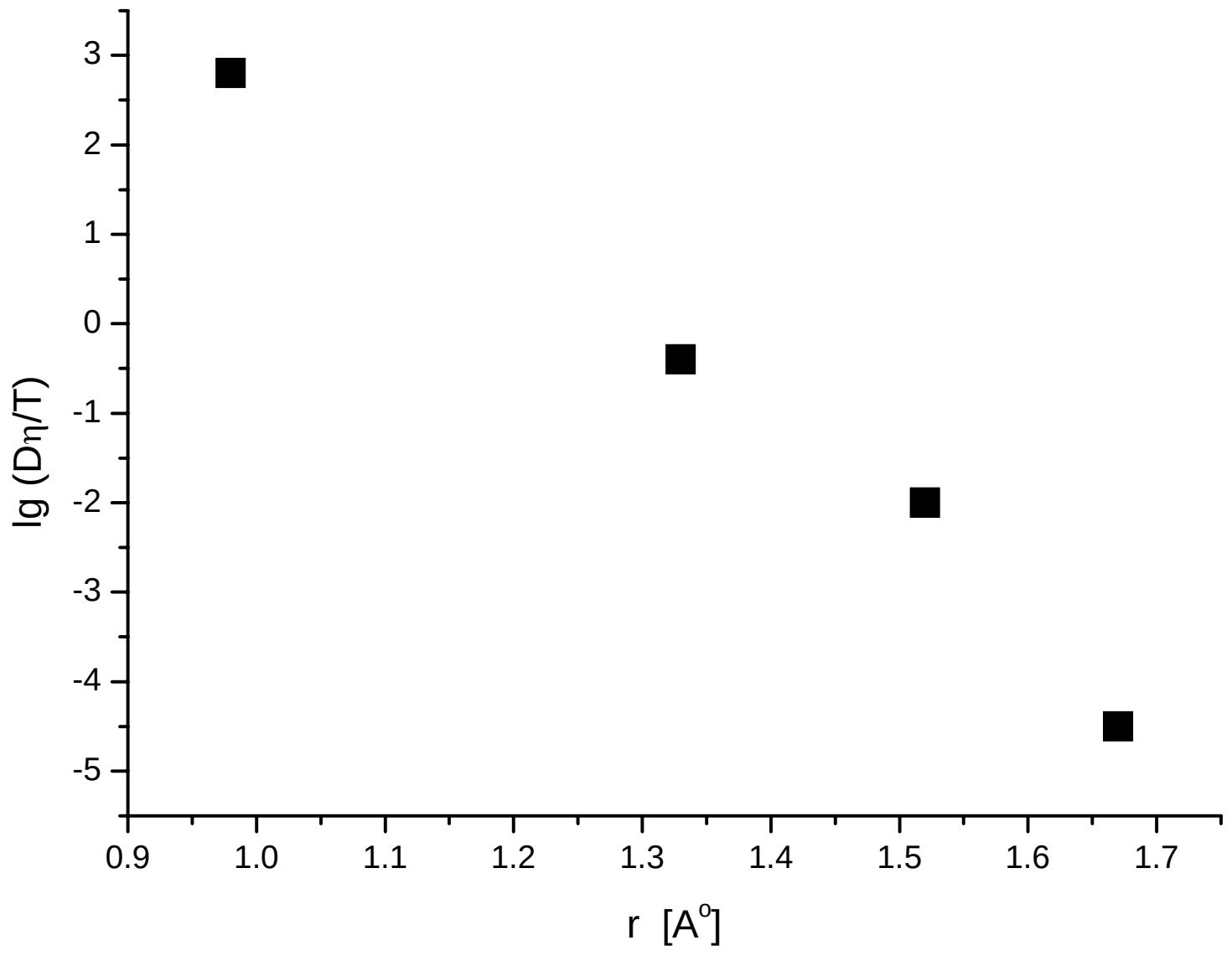

***Fig.4.*** *Dependence of the product lg(Dη/T) on the radius of the particles according to experimental data in Refs.[30,32-34].*

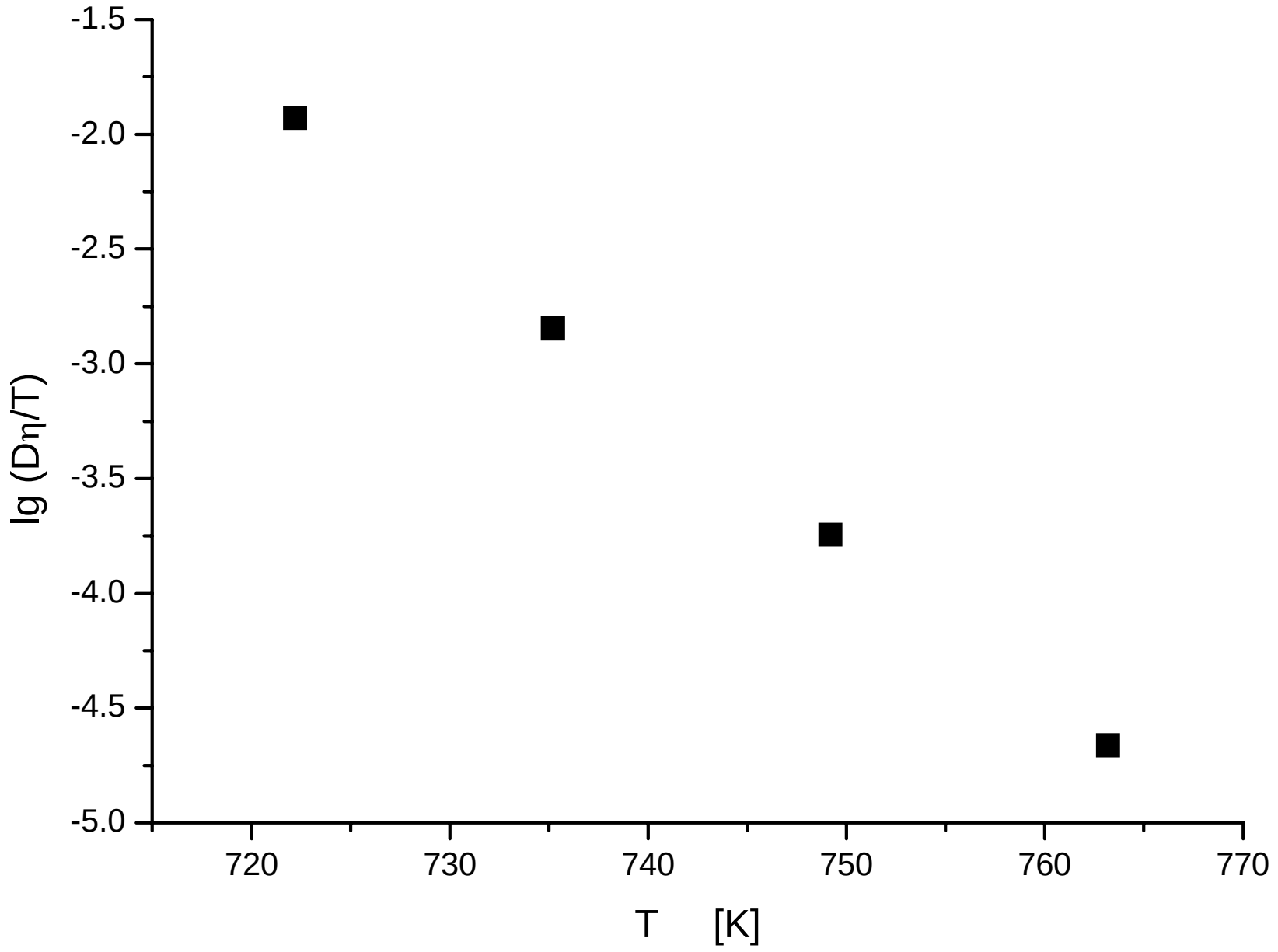

**Fig.5.** Dependence of the product *lg(Dη/T)* on the temperature for the experimental data in Ref.[30].

## 3. Conclusions

It is demonstrated that diffusion of small ions and atoms in glassforming systems proceeds according to a mechanism, essentially different from the mechanism of viscous flow. The smallest particles for which the Stokes-Einstein equation is valid are of the size of the main building units of the glassforming system. Smaller particles diffuse according to a mechanism discussed here.

As Einstein says: "A*ny physical theory cannot be fully testified but just disproved – we are often blind trying to understand our nature"*.